\documentstyle{mn}
\input epsf

\title[Frequency resolved spectroscopy of Cyg X-1]
{Frequency resolved spectroscopy of Cyg X-1: fast variability of
the reflected emission in the soft state.}

\author[M. Gilfanov, E.Churazov and M. Revnivtsev]
{M. Gilfanov$^{1,2}$, E.Churazov$^{1,2}$, M. Revnivtsev$^{2,1}$\\
$^1$Max-Planck-Institute f\"ur Astrophysik, 
Karl-Schwarzschild-Str. 1, 85740 Garching bei M\"unchen, Germany \\
$^2$Space Research Institute, Russian Academy of Sciences,
Profsoyuznaya 84/32, 117810 Moscow, Russia,\\
}

\begin{document}

\maketitle

\label{firstpage}

\begin{abstract}

Using the RXTE/PCA data we study the fast variability of the reflected
emission in the soft spectral state of Cyg X-1 by means of Fourier
frequency resolved spectroscopy. We find that the rms amplitude of
variations of the reflected emission has the same frequency dependence as
the primary  radiation down to time scales of $\la 30-50$ msec.  This
might indicate that the reflected flux reproduces, with 
nearly flat response, variations of the primary emission.
Such behavior differs notably from the hard spectral state, in which
variations  of the reflected flux are significantly suppressed in
comparison with the primary emission, on time scales shorter than
$\sim 0.5-1$ sec. 

If related to the finite light crossing time of the reflector, these results
suggest that the characteristic size of the reflector -- presumably an
optically thick accretion disk, in the hard spectral state is larger
by a factor of $\ga 5-10$ than in  the soft spectral state. 
Modeling the transfer function of the disk, we estimate the 
inner radius of the accretion disk $R_{in}\sim 100 R_g$ in the hard and 
$R_{in}\la 10R_g$ in the soft state for a $10M_{\sun}$ black hole. 
\end{abstract}
\begin{keywords}
accretion, accretion disks -- black hole physics --
stars: binaries: general -- stars: individual (Cygnus X-1) --
X-rays: general -- X-rays: stars
\end{keywords}

\section{Introduction}

The importance of the reprocessed/reflected component in the X--ray
spectra of accreting X--ray sources for exploring the geometry of the
accretion flow is well known. Reflection of the primary Comptonized
radiation from neutral or partially ionized matter located in 
the vicinity of the compact object -- presumably the optically thick
accretion disk, leads to appearance of characteristic
features in the spectra of X--ray binaries.  
The main signatures of the emission reflected from cold neutral
medium are well known -- the fluorescent K$_{\alpha}$ line of iron at 6.4
keV, iron K-edge at 7.1 keV  and a broad hump at $\sim 20-30$ keV
(Basko, Sunyaev \& Titarchuk 1974, George \& Fabian 1991).
The exact shape of these spectral features in the X-ray binaries depends on
ionization state of the reflecting medium and might be modified by strong
gravity effects and intrinsic motions in the reflector (e.g. Fabian et al.,
1989). The amplitude of the reflection signatures depends primarily on the
ionization state and the solid angle subtended by the reflector as seen from
the source of the primary radiation. 

Recently, Revnivtsev, Gilfanov \& Churazov (1999, hereafter Paper I)
proposed Fourier frequency resolved spectral analysis to 
study spectral variability in the X-ray binaries.
Although interpretation of the Fourier frequency resolved spectra in general
is not straightforward and requires  a priori assumptions to be made, one
of the areas where this method can be efficiently used is the fast 
variability of the reflected component. It has been found that in the low
spectral state of Cyg X-1 and GX339-4 the energy spectra corresponding to
the shorter time scales ($\la 0.1-1$ sec)  show less reflection than
those of longer time scales (Revnivtsev, Gilfanov \& Churazov
1999, 2000). The simplest, although not unique, interpretation of
this result is smearing of the short term variations of the reflected
emission due to finite light crossing time of the reflector. 
Based on the Fourier frequency dependence of the equivalent width of
the Fe K$_{\alpha}$ fluorescent line Revnivtsev et al. (1999) estimated the
characteristic size of the reflector:  $\sim  80-160 R_g$ for a $20-10
M_{\sun}$ black hole.

In this paper we investigate the fast variability of the reflected component
in the soft spectral state of Cyg X-1 and compare it with that in the
hard spectral state.

\section{Observations and data analysis.}

We used publicly available data of Cyg X--1 observations with the
Proportional Counter Array  aboard the Rossi X-ray Timing Explorer
\cite{rxte} performed in June, 1996 during the soft spectral state of the
source. The list of the observations is given in Table \ref{obslog}. The
total live time was $\approx 11.5$ ksec. The ``Generic Binned'' mode data in
configuration  B\_4ms\_8A\_0\_35\_H, with time resolution of $\approx 4$
msec and covering 2.9-13.1 keV energy range was used for frequency resolved
spectral analysis. 

The data screening and selection was performed using FTOOLS 4.2 with
standard screening criteria recommended by RXTE GOF. The
frequency resolved spectra were calculated following the prescription
detailed in Paper I. The dead time corrected value of the white noise level  
was determined from fitting of  the power spectra in the 300-1000 Hz
frequency range. The spectral analysis was performed in XSPEC v.10.0
\cite{xspec} with version 3.5 of the PCA response matrix. A uniform
systematic error of 0.5\% was added quadratically to the statistical
error in each energy channel.

The observations of the source 
during the hard state discussed in the text were performed between March 26
and 31, 1996. The details of these observations are given in Paper I.

\section{Results.}

\begin{figure}
\epsfxsize 9 cm
\epsffile{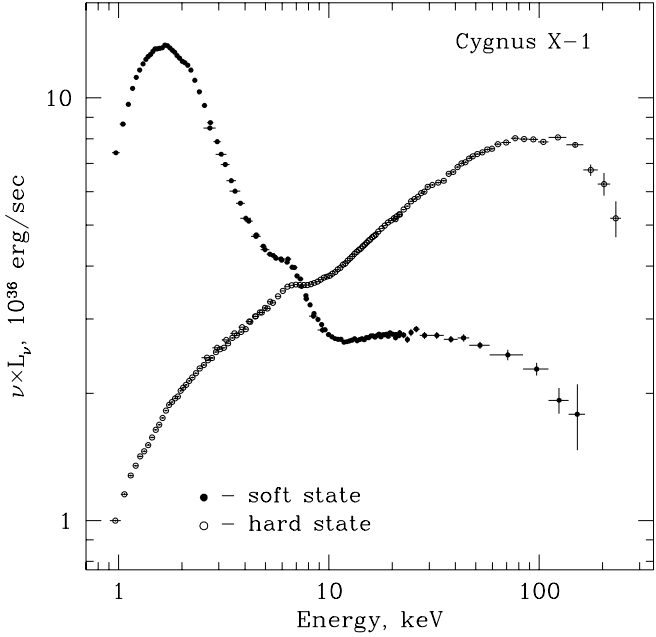}
\caption{ The spectral energy distribution of Cyg X--1 in the soft (filled
circles) and hard (open circles) spectral state. The data of nearly
simultaneous ASCA and RXTE observations on March 26, 1996 (hard state) and
May 30, 1996 (soft state). A source distance of 2 kpc was assumed
\protect\cite{zdz2}.} 
\label{spe}
\end{figure}

\begin{figure}
\epsfxsize 9 cm
\epsffile{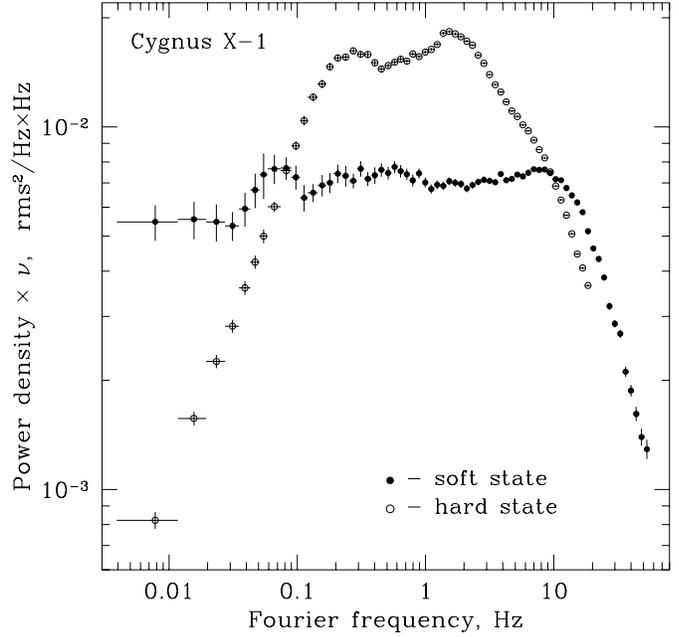}
\caption{ The power density spectra of Cyg X--1 in the soft (filled circles)
and hard (open circles) spectral state in the 3--13 keV energy band. 
The power spectra are plotted as (power density) $\times$
frequency, i.e. in units of $rms^2/Hz \times Hz$ and
are cut at approximately half of the Nyquist frequency (32
and 128 Hz for the hard and soft state respectively). No correction for
binning and aliasing effects has been applied.}
\label{pds}
\end{figure}

\begin{table}
\centering
\caption{The list of RXTE observations used for the analysis.
\label{obslog}} 
\begin{tabular}{ccccr}
\hline
Obs.ID & Date (UT) & Start & End & Expos.,s$^*$\\[10pt]
10512-01-05-00&	04/06/96&  20:49:36&  21:43:44&  1249\\
10512-01-07-00& 16/06/96&  00:07:44&  00:21:04&   643\\
10512-01-07-02& 16/06/96&  04:55:44&  05:17:04&  1113\\
10512-01-08-01& 17/06/96&  01:44:32&  01:56:00&   532\\
10512-01-08-02& 17/06/96&  04:56:32&  05:17:52&  1081\\
10512-01-08-00& 17/06/96&  08:08:32&  08:44:00&  1871\\
10512-01-09-02& 18/06/96&  03:21:36&  03:36:00&   675\\
10512-01-09-00& 18/06/96&  06:34:40&  07:01:52&  1461\\
10512-01-09-01& 18/06/96&  09:46:40&  10:26:56&  2170\\
\hline
\end{tabular}
\flushleft
$^*$ Dead time corrected PCA exposure time.
\end{table}

\begin{table*}
\centering 
\begin{minipage}{140mm}
\caption{Dependence of the best fit parameters upon Fourier frequency
\label{bestfit}} 
\begin{tabular}{ccccccc}
\hline
Frequency &  \multicolumn{3}{c}{\em soft state} & \multicolumn{3}{c}{\em hard state}\\
range, Hz & phot.ind. $\Gamma^1$ & EW$^2$, eV &EW/EW($<1$Hz)$^3$ & phot.ind. $\Gamma^1$ & EW$^2$, eV &EW/EW($<1$Hz)$^3$ \\
\hline

0.016--0.078&$2.60\pm 0.032$&$494\pm83$ &$1.05\pm 0.18$&$ 1.97\pm 0.006 $&$145\pm16.6$&$0.97\pm0.11$\\
0.078--0.20 &$2.60\pm 0.022$&$484\pm61$ &$1.03\pm 0.13$&$ 1.95\pm 0.004 $&$153\pm13.0$&$1.02\pm0.09$\\
0.20--0.45  &$2.63\pm 0.018$&$482\pm48$ &$1.02\pm 0.10$&$ 1.94\pm 0.004 $&$152\pm11.3$&$1.02\pm0.08$\\
0.45--1.0   &$2.62\pm 0.014$&$424\pm38$ &$0.90\pm 0.08$&$ 1.94\pm 0.004 $&$149\pm10.4$&$1.00\pm0.07$\\
1--2        &$2.63\pm 0.012$&$445\pm33$ &$0.94\pm 0.07$&$ 1.89\pm 0.004 $&$121\pm9.8$ &$0.81\pm0.07$\\
2--4        &$2.64\pm 0.011$&$429\pm31$ &$0.91\pm 0.07$&$ 1.86\pm 0.004 $&$121\pm9.9$ &$0.81\pm0.07$\\
4--8        &$2.64\pm 0.011$&$413\pm31$ &$0.88\pm 0.07$&$ 1.81\pm 0.004 $&$107\pm11.4$&$0.71\pm0.08$\\
8--16       &$2.58\pm 0.011$&$462\pm35$ &$0.98\pm 0.07$&$ 1.70\pm 0.007 $&$ 75\pm16.6$&$0.50\pm0.11$\\
16--32      &$2.52\pm 0.018$&$410\pm59$ &$0.87\pm 0.13$&$ 1.69\pm 0.017 $&$ 47\pm37.8$&$0.31\pm0.25$\\
32--128     &$2.43\pm 0.050$&$233\pm172$&$0.49\pm 0.37$&$--$&$--$&$--$\\
\hline
\end{tabular}\\
\flushleft
The errors are $1\sigma$ for one parameter of interest.\\
$^1$ -- the power law photon index;\\
$^2$ -- the equivalent width of the 6.4 keV line, eV;\\
$^3$ -- EW($<1$Hz) -- equivalent width averaged in the 0.016-1.0 Hz
frequency range
\end{minipage} 
\end{table*}

The broad band energy spectra of Cyg X-1 in the hard and soft spectral state
are shown in Fig.\ref{spe}. The spectra were obtained using the data of
overlapping ASCA and RXTE observations of the source on March 26, 1996 (hard
state) and on May 30, 1996 (beginning of the soft state).

The power spectra of Cyg X-1 in the 3--13 keV energy band in the  hard and
soft spectral states obtained from the complete sets of the data used for the
frequency resolved spectral 
analysis in this paper and in Paper I are shown in Fig.\ref{pds}. 
The power density is plotted in units of power$\times$frequency representing 
squared fractional rms at a given frequency per factor $\sim e$ in
frequency. This way of representing the power spectra most clearly
characterizes  relative contribution of variations at different
frequencies to the total observed rms.

The Fourier frequency resolved spectra for the soft spectral state were
obtained in 10 frequency 
bins of logarithmically equal width in the 0.016-128 Hz frequency range.  
In order to study the frequency dependence of the amplitude of the
reflected component we fit the spectra in the 3--13 keV band with a
simplified model consisting of an absorbed power law\footnote{Contrary
to the average spectra of Cyg X-1 in the soft state (Fig.\ref{spe}),
the contribution of the soft component to the frequency resolved
spectra is negligible (to be discussed in more detail in a separate
paper). Therefore use of a power law to model continuum emission in
the soft state is justified.} 
with superimposed Gaussian line at 6.4 keV.  The low energy absorption
was fixed at $N_H=6\cdot 10^{21}$ cm$^{-1}$. The line width (standard
deviation for a Gaussian profile) was fixed at 1 keV 
which corresponds to the average value for the high state observations used
for the analysis. The power law photon index and line flux were the only
free parameters of the fit. To facilitate comparison with the hard state
data we reanalyzed the data set used in Paper I in identical
frequency bins covering 0.016-32 Hz frequency range (the time resolution of
the hard state data was $\approx 16$ msec) and using the same spectral
model. The width of the line for the hard state spectra was fixed at the
corresponding average value of 0.8 keV \cite{cygx1}.

Such a spectral model is obviously oversimplified. Neither it is  justified
from the physical point of view. It is, however, suitable to quantify the
amplitude of the characteristic ``wiggle'' usually seen in the spectra of the
accreting X--ray sources in the $\sim 5-15$ keV energy range and commonly
attributed to the effects of reflection. Use of more sophisticated
models is restricted by insufficient statistics (especially in the
soft spectral state) and the low number of energy channels below
$\approx 13$ keV in the B\_4ms\_8A\_0\_35\_H configuration used in the
most of the soft state observations.

\begin{figure}
\epsfxsize 9 cm
\epsffile{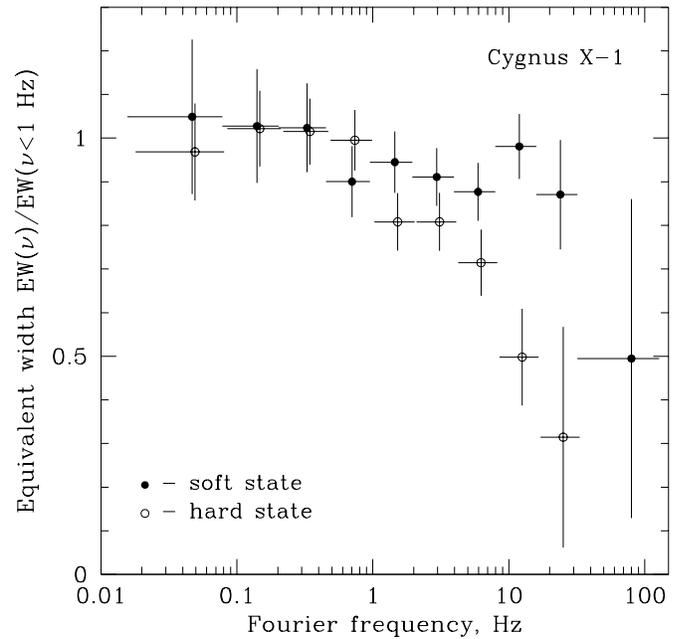}
\caption{Equivalent width of the Fe fluorescent line vs. Fourier frequency in
the soft (filled circles) and hard (open circles) spectral state. The
equivalent width is normalized to the 0.01-1 Hz average.}
\label{eqw}
\end{figure}

The best fit values of the spectral parameters for both spectral states are
given in Table \ref{bestfit}. The dependence of the equivalent width
on the Fourier frequency is shown in Fig.\ref{eqw}. 
Variations of the parameters of the spectral model, in particular the change
of the line centroid from 6.0 to 6.7 keV  and reducing the intrinsic line
width to zero, do not change the general trend. These variations, however,
affect the particular values of the equivalent width and, to lesser extent,
the shape of the curve in Fig. \ref{eqw}.

According to the $\chi^2$--test two distributions, shown in
Fig.\ref{eqw} differ at the confidence of level of $\approx 98.7\%$
($\chi^2=19.4$ for 8 d.o.f. in the 0.016--32 Hz frequency range). 
It should be noted, however, that the errorbars assigned to the Fourier
frequency resolved spectra were propagated  from the corresponding power
density spectra and are likely to be somewhat overestimated, especially
in the low frequency bins (this fact can be noticed in Fig.\ref{eqw}).
The confidence level, calculated using 0.45--32 Hz
frequency range is $\approx 99.7\%$ ($\chi^2=18.1$ for 5 d.o.f.).

\section{Frequency dependence of the equivalent width and time response of
the reflector.}

\begin{figure*}
\hbox{
\epsfxsize 9 cm
\epsffile{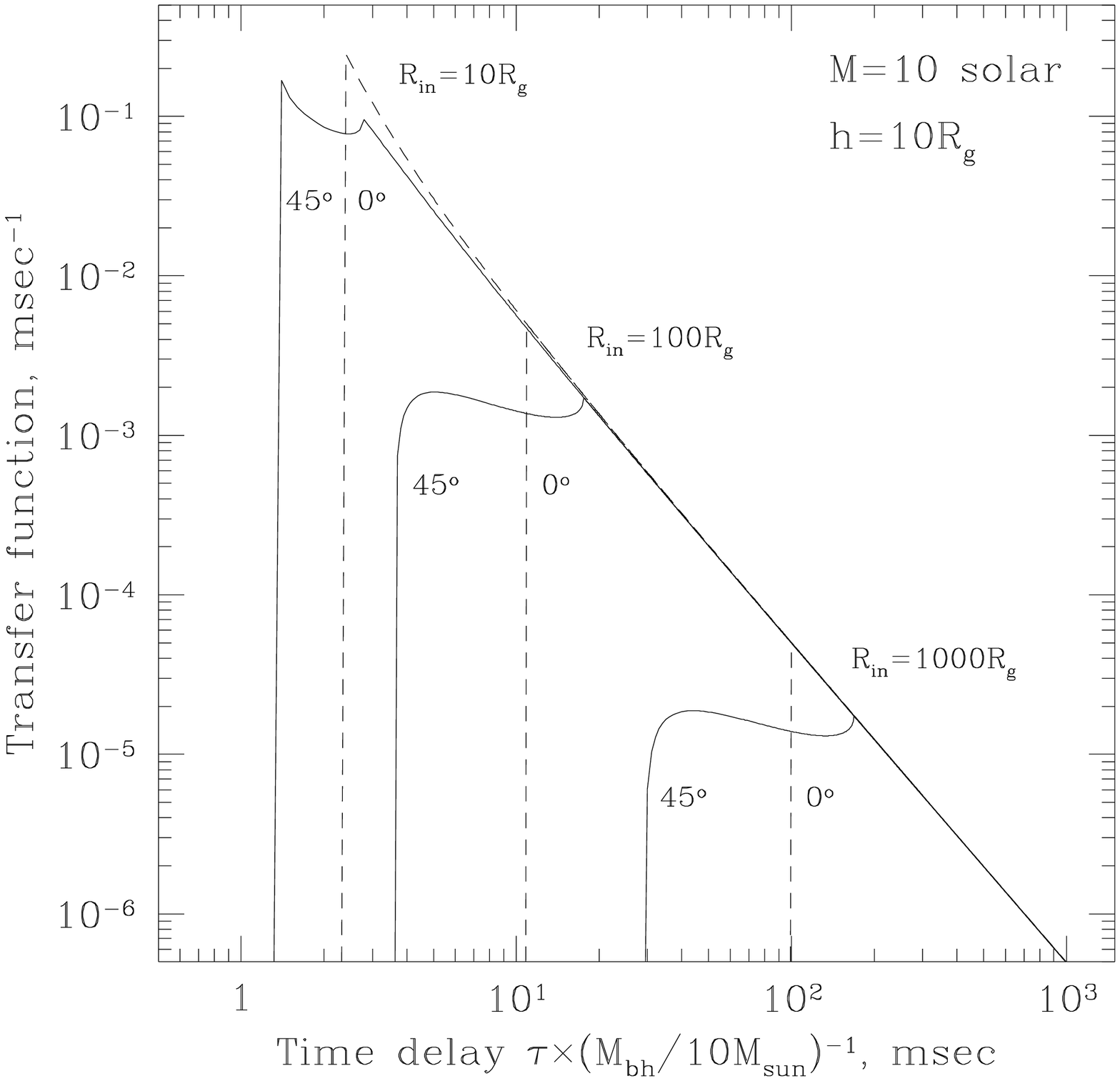}
\epsfxsize 9 cm
\epsffile{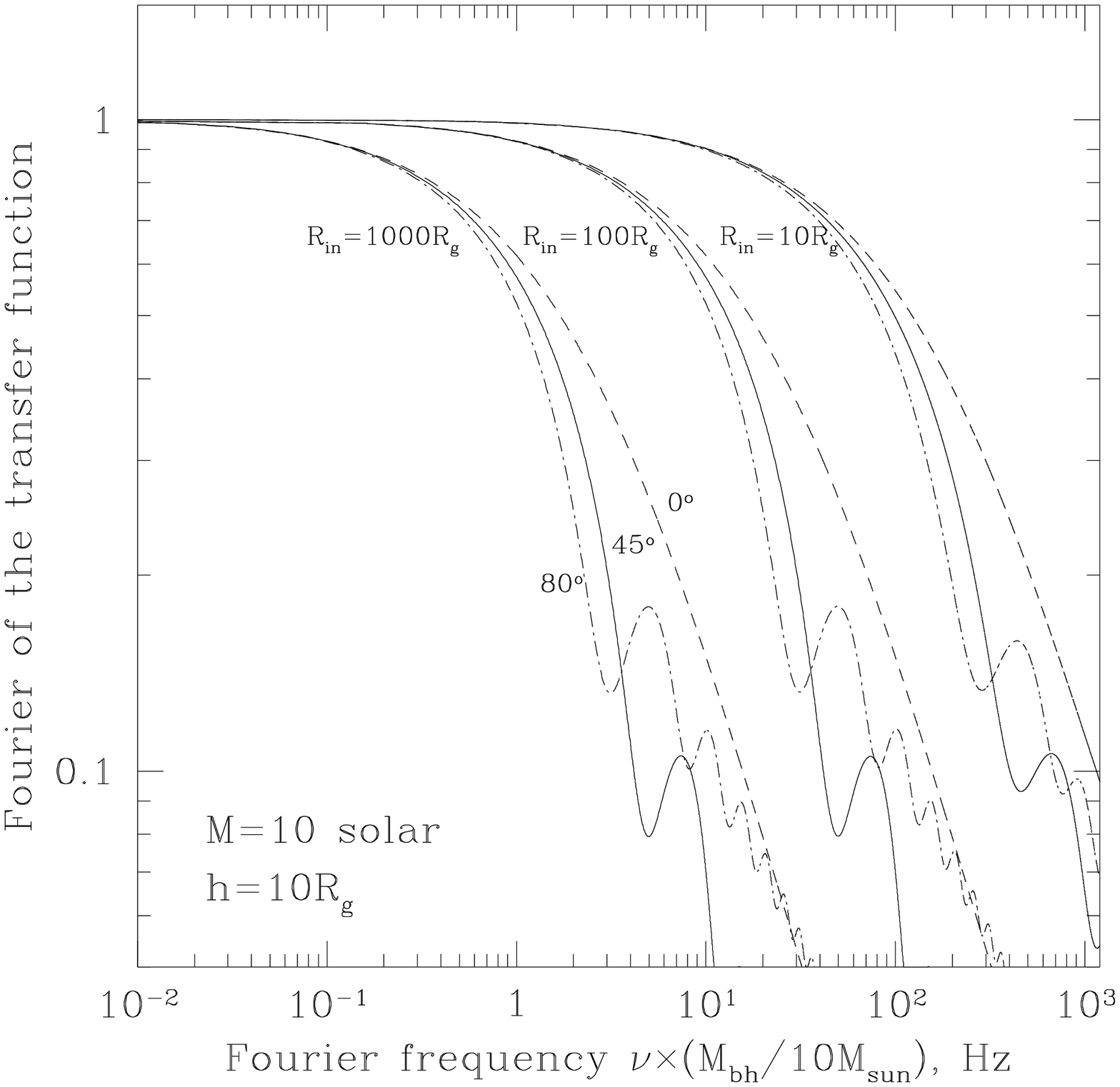}
}
\caption{Transfer function ({\em left}) of the disk and its Fourier
transform ({\em right}) for an isotropic point source of primary radiation
located at $h=10R_g$ above a flat disk with inner radius of
$R_{in}=10,100$ and $1000 R_g$ and inclination angle of $0\degr, 45\degr$
and $80\degr$ (not shown in the left panel). No relativistic effects have
been taken into account. The linear quantities are given 
in the units of gravitational radii for a $10M_{\sun}$ black hole.}
\label{timeresp}
\end{figure*}

\begin{figure*}
\hbox{
\epsfxsize 9 cm
\epsffile{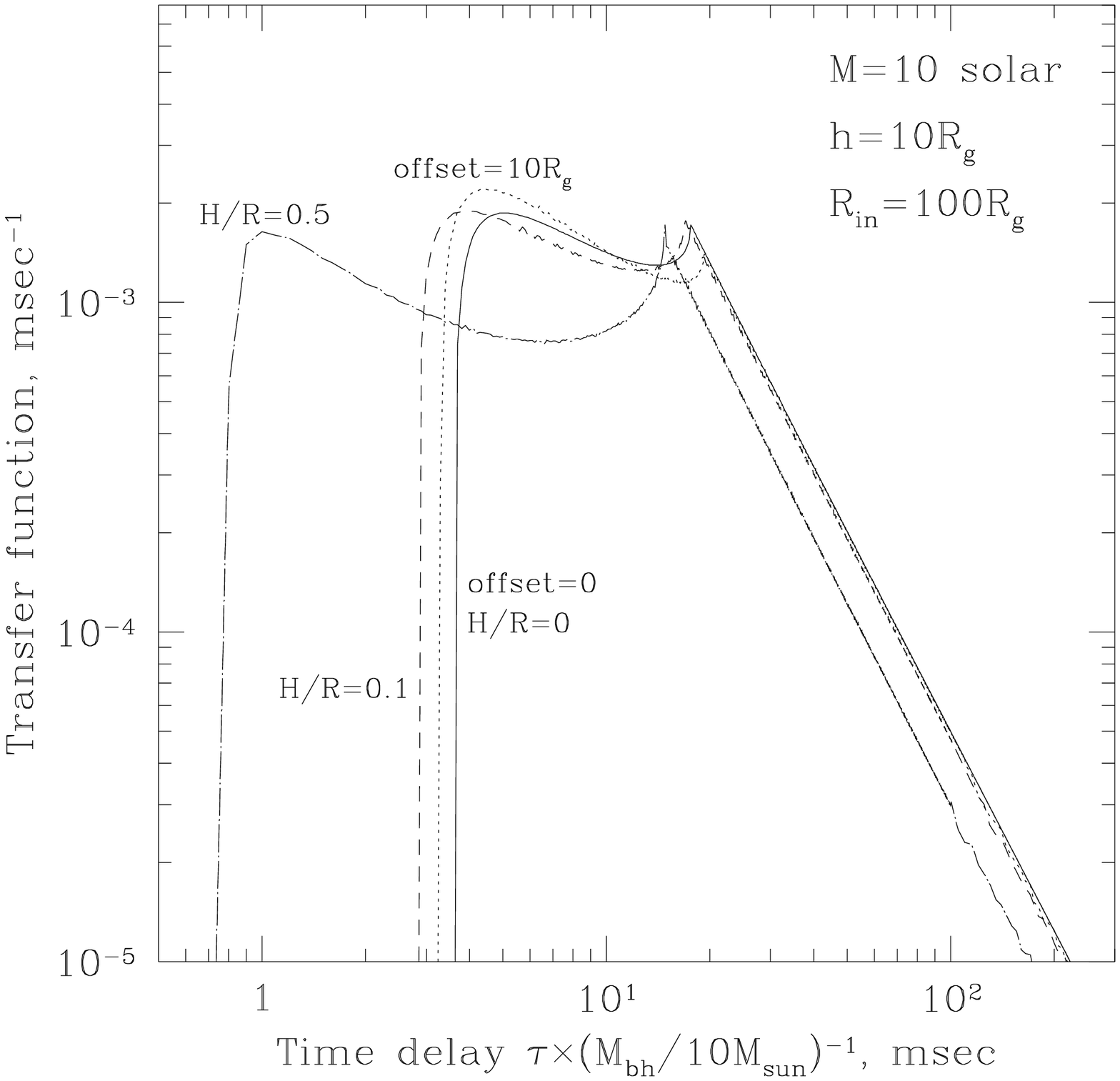}
\epsfxsize 9 cm
\epsffile{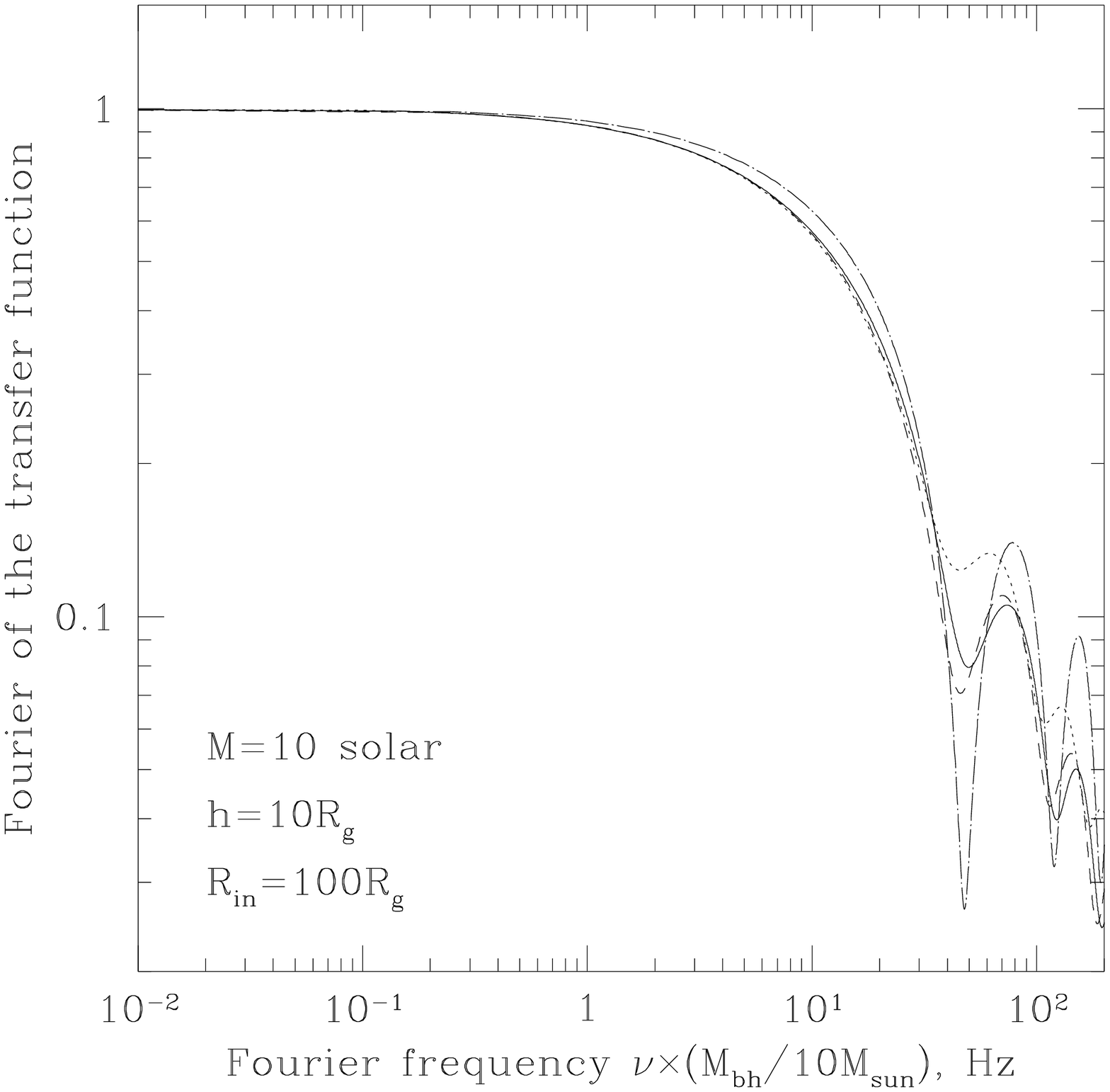}
}
\caption{Dependence of the transfer function and its Fourier transform upon
offset of the source of the primary radiation from the disk axis
(offset=$10R_g$, flat disk -- dotted line) and opening angle of the disk
(offset=0, $H/R=0.1$ -- dashed line and $H/R=0.5$ -- dash-dotted line).
Other parameters of the model are: inner disk radius $R_{in}=100 R_g$,
$M=10M_{\sun}$, inclination $45\degr$. No relativistic effects have
been taken into account.}
\label{ri100var}
\end{figure*}

The geometry, element abundances and ionization state of the reflector
being fixed, the equivalent width of the Fe fluorescent line
determined from a conventional energy spectrum is  proportional to the
relative amplitude of the reflected component and approximately
measures the solid angle subtended by the reflector. The equivalent
width of the fluorescent line determined from a Fourier frequency
resolved spectrum measures the ratio of the rms amplitudes of
variations of the reflected component and the primary emission  in a
given Fourier frequency range.  The constancy of the equivalent width of
the line at Fourier frequencies $f\la30$ Hz observed in the soft  
state (Fig.\ref{eqw}) implies that the rms amplitude of variations of
the reflected component has the same frequency dependence as that of the 
primary radiation. This would naturally appear if the reflected emission was
reproducing variations of the primary radiation down to the time scales of
$\sim 30-50$ msec. Such behavior is in contrast to the hard spectral state,
in which variations of the reflected flux are notably suppressed in
comparison with the primary emission on the time scales  shorter than $\sim
500$ msec.

The most straightforward explanation of this effect  would be in
terms of a finite light crossing time of the reflector 
$\tau_{\rm refl}\sim l_{\rm refl}/c$ 
due to its finite spatial  extend $l_{\rm refl}$. 
In this case the frequency dependence of the fluorescent line equivalent
width, $EW(f)$, is determined by the geometry of the primary source and the 
reflector. The characteristic width of the $EW(f)$ is mainly defined by the
spatial extent of the part of the accretion disk giving the main 
contribution to the reflected emission. 
Based on these arguments Revnivtsev, Gilfanov \& Churazov (1999)
estimated the characteristic size of the reflector in the low spectral
state as $l_{\rm refl}\sim 5\times 10^8$ cm which would correspond to
$\sim 150 R_{\rm g}$ for a $10M_{\sun}$ black hole.

Below we consider this problem in a more quantitative way. The time
dependence of the reflected emission is defined by the following relation: 
$$
F_{refl}(t)=\int\limits_{0}^{\infty} F_0(t-\tau) T(\tau) d\tau
$$
where $F_0(t)$ and $F_{refl}(t)$ are primary and reflected flux, $T(\tau)$
-- the transfer function of the reflector, defined by the geometry. This
function accounts for the propagation time of the photons from the
primary source to different parts of the reflector and then to the observer. 
The Fourier transform of the reflected flux is:
$$
\hat F_{refl}(f)=\hat F_0(f)\times \hat T(f)
$$
where $\hat F_0$, $\hat F_{refl}$ and $\hat T$ are corresponding Fourier
transforms and $f$ is Fourier frequency. The equivalent width of the
fluorescent line determined from the Fourier frequency resolved spectra is  
$$
EW(f)\propto\frac{|\hat F_{refl}(f)|}{|\hat F_0(f)|}=|\hat T(f)|
$$
i.e. is proportional to the Fourier transform of the transfer function of the
reflector.

Fig.\ref{timeresp} shows the transfer function and its Fourier transform for
an isotropic point source located at the height $h$ above a flat disk with
the inner radius $R_{in}$ and inclination angle $i$ for
different values of $R_{in}$ and $i$. A Lambert law for the angular
dependence of the reflected flux has been assumed. No general or  
special relativity effects have been taken into account. 
The characteristic width of the $EW(f)$ dependence is mainly defined by the
distance $d=\sqrt{h^2+R_{in}^2}$ 
between the primary source and the inner edge of the disk and depends
only weakly  on the inclination angle $i$. A small offset of the primary
source, $\Delta\ll d$ or non-zero opening angle of the disk do
not significantly affect the characteristic width of Fourier
transform of the response function  (Fig.\ref{ri100var}). However, the
transfer function itself and the high frequency part of its Fourier
transform are sensitive to the details of the geometry
(Fig. \ref{timeresp} and \ref{ri100var}). 

\begin{figure}
\epsfxsize 9 cm
\epsffile{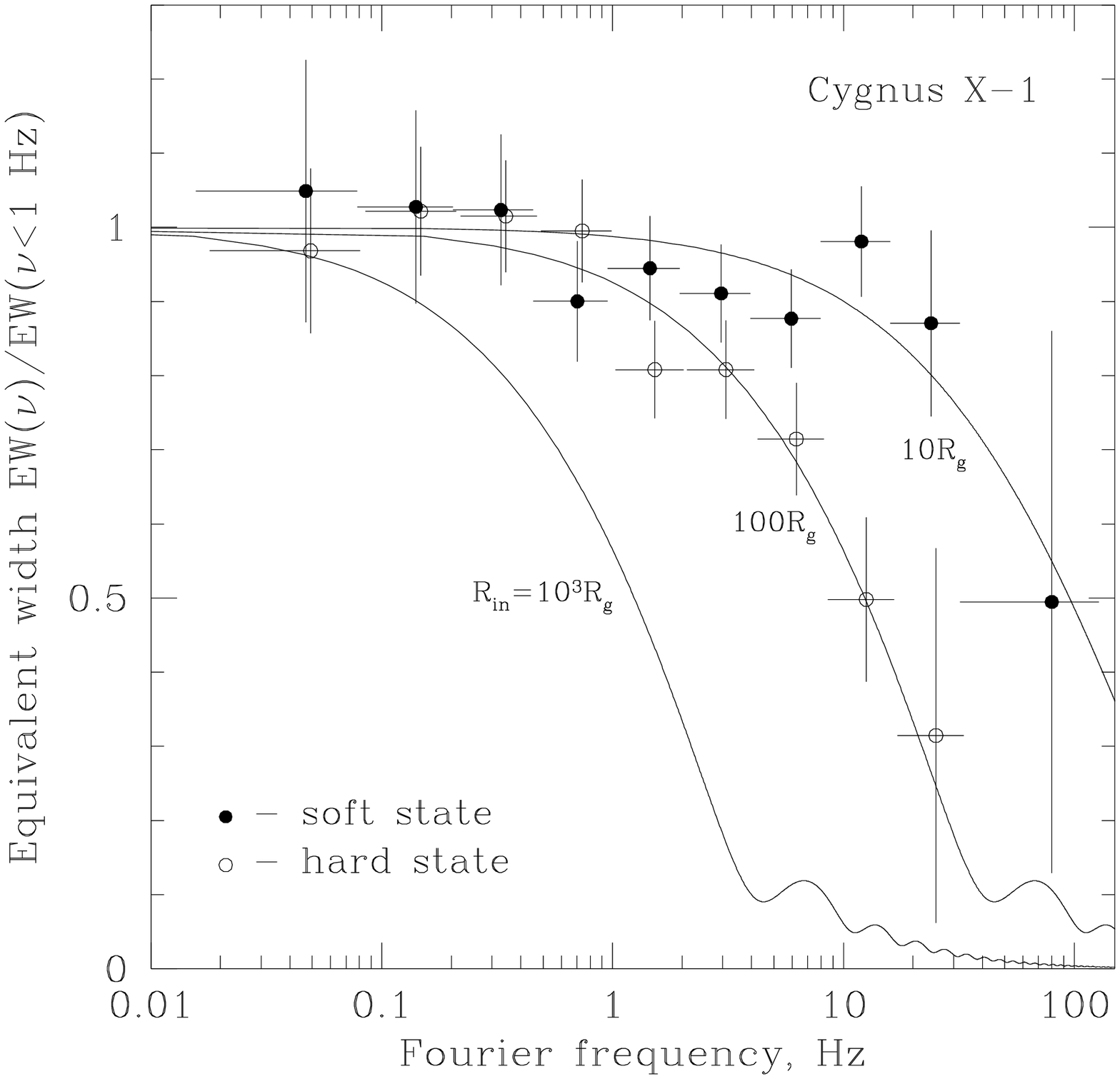}
\caption{Equivalent width of the Fe fluorescent line vs. Fourier frequency.
Comparison of the data with the model. The data points are the same as in
Fig.\ref{eqw}. The model curves were calculated for an isotropic point
source at height $h=10R_g$ on the axis of a flat disk with inner radius of
$10,100$ and $1000 R_g$ (assuming a $10M_{\sun}$ black hole) and with
inclination angle of $50\degr$.}
\label{eqw_mod}

\end{figure}

In Fig.\ref{eqw_mod} we compare Fourier transform of the transfer function of
a flat disk with inclination $i=50\degr$, appropriate for Cyg X-1
\cite{incl1}, with the frequency dependence of the equivalent width observed
in the soft and hard spectral states of Cyg X-1. 
As seen from Fig.\ref{eqw_mod}, suppression of the high frequency variations 
in the reflected emission observed in the hard state can be satisfactorily 
described by reflection from a disk with an inner radius of $R_{in}\sim 100
R_g$ around a $10M_{\sun}$ black hole. Significantly larger, $R_{in}\sim 1000
R_g$, or smaller values of the inner radius, $R_{in}\sim 10 R_g$, are
inconsistent with the data. The soft state data, on the other hand,
requires much smaller values of the inner radius of the disk,
$R_{in}\sim 10R_g$.

It should be noted that since we use the equivalent width of the Fe
fluorescent line as a measure of the amplitude of the reflected emission,
the results might be somewhat affected by the non-uniformity of the
ionization state of the accretion disk with radius and geometrical effects
(e.g. radial and azimuthal dependence of the reflection angle).  
Results of more detailed modeling of the disk transfer function and
rigorous comparison with the data will be published elsewhere
(a paper in preparation).

\section{Discussion.}

Study of fast variability of the reflected emission by means of Fourier
frequency resolved spectroscopy offers a new independent method to probe the
geometry of the accretion flow which complements conventional X-ray
spectral analysis. The presence of a luminous soft component dominating the
X-ray spectrum in the soft spectral state of black hole candidates
(Fig.\ref{spe}) strongly favors small values of the inner radius of the
disk (e.g. Gierlinski et al., 1999), in good agreement with the above
result, $R_{in}\la 10R_g$. 
In the hard spectral state  conventional estimates of the inner radius of
the accretion disk range from several tens $R_g$ in coronal model
(e.g. Poutanen, Krolik \& Ryde, 1997; Done \& Zicky, 1999) to several
hundred $R_g$ in ADAF model  (e.g. Esin et al., 1998). Our result,
$R_{in}\sim 100 R_g$, falls in the middle of this  range.    

The inner radius of the accretion disk determined in the above analysis
refers to the inner radius of the ``reflective'' part of the disk, where
the ionization state is such that the disk is capable to produce a
fluorescent iron line. Therefore substantial change of the ionization
state of the surface layer of the inner disk (e.g. Young et al., 1999)
may have similar effect on the frequency dependence of the equivalent width
of the iron line as physical change of the inner disk radius.

In the simplified geometry of an isotropic point source above flat disk
solid angle subtended by the reflector is  
$\frac{\Omega}{2\pi}=\frac{1}{\sqrt{1+R_{in}^2/h^2}}$ (in the notation of
the previous section). For the parameters from Fig.\ref{eqw_mod}
respective values of the solid angle are: 
$\frac{\Omega}{2\pi}\approx 0.1$ for the hard state 
($h=10R_g$, $R_{in}=100R_g$) and  
$\frac{\Omega}{2\pi}\approx 0.7$ for the soft state 
($h=10R_g$, $R_{in}=10R_g$). Equivalent width of the iron fluorescent line
expected for these values of the solid angle and solar abundance of
iron are several times smaller than those given in the Table
\ref{bestfit}, especially for the hard spectral state. However, as was
noted above, the absolute values of the equivalent 
width quoted in Table \ref{bestfit} are subject to some uncertainty due to 
simplified model used for the spectral fits.  Comparison
of the model predictions with more accurately determined values of the
equivalent width and amplitude of the reflected component can further
constrain geometry of the accretion flow.

Finally, we should note that interpretation of the frequency dependence of
the equivalent width of the fluorescent line in  terms of the finite light
crossing time of the reflector is not unique.   
An alternative explanation might be that the short time scale, $\la$
50--100 msec, variations appear in a geometrically different, likely
inner, part of the accretion flow and give a rise to significantly
weaker, if any, reflected emission than the longer time scale events
presumably originating in the outer regions. This might be caused, for
instance,  by a smaller solid angle of the reflector as seen by the short
time scale events and/or due to screening of the reflector from  the short
time scale events by the outer parts of the accretion flow. 
However, independent of the nature of the fall off of the equivalent width 
at high frequency, the flat response of the reflected emission to variations
of the primary radiation observed at low frequencies puts an upper limit  on
the spatial extent of the reflector, i.e. on the inner radius of the
accretion disk. In particular, large values of $R_{in}$, significantly
exceeding $\sim 100 R_g$ in the hard and $\sim 10R_g$ in the soft spectral
state are  excluded by our analysis.

\section{Conclusions.}

We have exploited Fourier frequency--resolved spectral analysis to
study fast variability of the reflected emission on time scales of 
$\sim 100$ sec -- 10 msec in the soft and hard spectral states of Cyg X-1. 
Our conclusions are:
\begin{enumerate}
\item
In the soft spectral state variations of the reflected component have the
same frequency dependence of the rms amplitude as the primary emission 
up to the frequencies $\ga 30$ Hz.
This would be expected if, for instance, the reflected flux was
reproducing, with flat response, variations of the primary radiation
down to the time scales of $\sim 30-50$  msec. The sensitivity of the
present analysis is insufficient to study shorter time scales. 
\item
In the hard spectral state variability of the reflected flux is
significantly suppressed in comparison with the direct emission
on the time scales shorter than $\sim 0.5-1$ sec (see also Paper I). 
\item
Assuming that suppression of the short-term variability of the reflected
emission is caused by the finite light-crossing time of the reflector, we 
estimated the inner radius of the accretion disk $R_{in}\sim 100 R_g$ in
the hard spectral state and $R_{in}\la 10 R_g$ in the soft spectral state. 
\end{enumerate}

\section*{acknowledgments}
This research has made use of data obtained through the High Energy
Astrophysics Science Archive Research Center Online Service, provided 
by the NASA/Goddard Space Flight Center. 
The work was done in the context of the research network
"Accretion onto black holes, compact objects and protostars"
(TMR Grant ERB-FMRX-CT98-0195 of the European Commission).
M.Revnivtsev acknowledges partial support by RBRF grant 97-02-16264
and INTAS grant 93--3364--exit.

\bsp

\label{lastpage}

\end{document}